# Toward a Distributed Knowledge Discovery system for Grid systems

Nhien An Le Khac[1] , Lamine M. Aouad[2] , and M-Tahar Kechadi[3]

[1] School of Computer Science & Informatics, University College Dublin
    `an.lekhac@ucd.ie`
[2] School of Computer Science & Informatics, University College Dublin
    `lamine.aouad@ucd.ie`
[3] School of Computer Science & Informatics, University College Dublin
    `tahar.kechadi@ucd.ie`

## 1 Introduction

During the last decade or so, we have had a deluge of data from not only science fields but also industry and commerce fields. Although the amount of data available to us is constantly increasing, our ability to process it becomes more and more difficult. Efficient discovery of useful knowledge from these datasets is therefore becoming a challenge and a massive economic need. This led to the need of developing large-scale data mining (*DM*) techniques to deal with these huge datasets either from science or economic applications.

Moreover, these large volumes of data that are collected daily are often heterogeneous, geographically distributed and owned by different organisations. In a distributed environment, datasets are distributed among different sites for various reasons. For example, an application by its nature is distributed such as a multinational company that has customers worldwide - the datasets concerning its products and customers are distributed and heterogeneous (different customers, different legislation from country to country, etc.). In scientific applications, for instance, the data may be collected in different locations using different instruments, and therefore these separate datasets may have different formats and features. Concretely, the datasets we produce today are by nature distributed both in content and in administrative policies. What is perceived by the end user as a single data collection (dataset or database or data warehouse) is in fact composed of different collections of data administered by different authorities and owned by a variety of organisations with diverse or competing business models and strategies. Traditional centralised data management and mining techniques are not adequate anymore. More precisely, traditional *DM* techniques do not consider all the issues of data-driven applications such as scalability in both response time and accuracy of solutions, distribution and heterogeneity. In addition, transferring a huge amount of data  over the network  is not an efficient strategy and may not be possible for security and protection reasons.  For example,  in some fields such as physics (e.g., LHC at CERN,  nanotechnology), bioinformatics (predicting gene, mapping/sequencing DNA, etc.), digital business ecosystem, meteorology, digital  forensic, and telecommunication, where the data  is produced and stored locally (distributed data) the users (scientists or data  analysts) need to unite their effort to mine and analyse the data.  At the same time they need to share the access to the data  of interest. This puts a huge stress on the data integrity and protection for following main reasons: (i) different organisations have different policies for the access rights; (ii) the current Internet and other Global computers  are very vulnerable  to attacks, such as viruses, hacking, and Denial of Service attacks; (iii) network  failure and temporarily heavy traffic.

Distributed data  mining techniques  constitute a better alternative as they are scalable and can deal efficiently with data  heterogeneity. So distributed data  mining (DDM) has become  necessary  for large and multi-scenario datasets requiring  resources, which are heterogeneous and distributed. There are two major  strands of research into DDM. The first strand considers homogeneous data  sites and consists of combining different models of data  from different sites.  The second strand considers a broader range of distribution, where the datasets located in different sites may record different features. Most research effort is concentrated on the first strand and several techniques can be found in the literature. There exists very little that addresses the second strand (DDM on heterogeneous  datasets). Moreover, existing methods  are limited, particularly when additional requirements are needed and complexity (distribution, heterogeneity, large volume of data) of real-world  data-driven applications  increases.

Besides, to cope with large, graphically  distributed, high dimensional, multi-owner, and heterogeneous  datasets, Grid platforms  [32] are well suited for data  storage and they provide  an effective computational support for distributed data  mining applications. Because recent Grid platforms,  which benefits from Web Services through its Open Grid Service Architecture [35] and Web Service Resource Framework  [23], is an integrated infrastructure that efficiently supports the sharing and coordinated use of resources in dynamic heterogeneous distributed environments. Actually,  only few projects that consider Grid as a platform  for distributed data  mining have been initiated [15, 9, 22] so far.  Most of them use some basic Grid services and they are based on Globus Tool Kit [36]. However, they only provide  means of managing and controlling  the resources of the Grid but they are not focused on how to take advantage of the  data-driven application features  to efficiently execute their corresponding  distributed algorithms.

In this chapter,  we present a new *DDM* system  combining dataset-driven and architecture-driven strategies. Data-driven  strategies will consider  the size and heterogeneity of the data,  while architecture driven will focus on the distribution of the datasets. This system is based on a Grid middleware tools that integrate appropriate large data  manipulation operations. Therefore,  this allows more dynamicity and autonomicity during the mining, integrating and processing phases. The following section presents issues related to a *DDM* system. Section 3 deals with our system architecture. Section 4 presents new *DDM* algorithms, the core of this system. One of its key layers, Knowledge Map (KM), will be described in Section 5; and then the exploitation of this system will be showed in the Section 6. We resumes related works of *DDM* systems on Grid platforms in Section 7. Finally, we conclude on Section 8.

## 2 *DDM* systems

In this section we firstly resume different aspects of *DDM* systems and then we discuss on problems related to an efficient *DDM* system on Grid platform.

A *DDM* system normally includes main components such as: data pre-processing, mining algorithms, communication subsystem, resource and task management, user interface. The main role of a *DDM* system is to provide an environment for accessing distributed data, mining algorithms and computing resource, monitoring the entire mining process, interpreting results to users. A *DDM* system should offer a flexible environment to adapt various kind of distributed mining applications. The architecture of a *DDM* system is also a important issue. Early researches on this subject are based on cluster of high-performance workstations or three-tier client/server model [18]. However, these approaches are appropriate for *PDM* tasks. Another approach is based on agent-based model addressing to scalable mining over large distributed datasets. Most of these systems require a supervisory agent that handles and facilitates the mining process.

### 2.1 *DDM* issues

**Centralised DM vs. *DDM***

Today, traditional centralised DM is not suitable for exploring a huge amount of data distributed in large scale environments. Some of the main problems can be listed as the communication cost, bottleneck, using of distributed resources, privacy and security, etc. In a distributed environment, data may be distributed among different sites for various reasons: an application by its nature is distributed or sometimes data are artificially distributed for better scalability and disk space management. Centralised DM techniques do not consider all the issues of data-driven applications such as scalability in both response time and accuracy of solutions, distribution and heterogeneity. Meanwhile, *DDM* approaches, as shown in the Figure **??**, perform local data analysis followed by the generation of a global model by aggregating the local results. Precisely, *DDM* techniques which base on the availability of the distributed resources can be able to learn models from distributed data without exchanging the raw data.

**Parallel Data Mining and** *DDM*

The objective of Parallel Data Mining (PDM) is to perform fast mining of large datasets by using high performance parallel environments. We can find some related works on PDM in the literature such as [3, 34, 44]. This approach assumes existence of high-speed network connection between the computing nodes. That is not always available in many of the *DDM* applications. In spite of the development of *DDM* has been influenced by PDM, this approach is not in the scope of this book chapter where we investigate *DDM* techniques scaling well on large environments without existing of high-speed network connection.

**Homogeneous vs. Heterogeneous**

As mentioned in the Section 1, there are two major strands of research into *DDM*. The first strand considers homogeneous data sites and consists of combining different models of data from different sites. The second strand considers a broader range of distribution, where datasets located in different sites may record different features. This strand is also called *DDM* on heterogeneous datasets. Most research effort is concentrated on the first strand and several techniques can be found in the literature. For instance, [8, 21, 54] proposed ensemble learning for distributed classifier learning. Meta-learning [17] offers another approach for learning classifiers from homogeneous distributed data. Distributed Association Rule mining and distributed clustering for this homogeneous case can be found in [1, 31].

There exists very little research in the literature that addresses to the second trend ( *DDM* on heterogeneous datasets). Most of them uses special-purpose algorithms. The WoRLD system [**?**], for example, is based on the "activation spread" approach. It first computes locally the cardinal distribution of the feature values of datasets. In the second phase, this knowledge is propagated across different sites. Features with strong correlations to the space model are identified and selected, to be use for learning the distribution. However, this technique may not be always appropriate for a given space model.

## 2.2 Toward an efficient *DDM* framework on Grid platform

Traditional approaches of *DM*, as resumed above, are limited, particularly when the additional requirements and complexity (distribution, heterogeneity, large volume of data) of real-world data-driven applications are included. These requirements constitute challenges in this research area. Today, the development of Grid technologies allows to share resources distributed in large, heterogeneous environments. However, the sharing and transferring of a huge amount of data is not efficient and sometimes is not impossible because of the performance aspects. *DDM* becomes a remarkable solution for mining applications distributed on Grid platform. There are many challenges concerning both *DDM* techniques and the infrastructure that allow efficient and fast processing, reliability, quality of service, integration, and extraction of knowledge from this mass of data.

In order to become an efficient platform to explore and analysis huge data distributed on the Grid environment, a *DDM* system needs to combine dataset-driven and architecture-driven strategies. Dataset-driven will take into account the size and heterogeneity of data while architecture-driven will take into account the distribution of datasets. Concretely, an efficient *DDM* framework is consisted of the following main features:

- Providing a set of *DDM* algorithms that are robust, adaptive, flexible, low cost communication and scalable. *DDM* algorithms are heart of *DDM* system. However, most of the current approach of *DDM* framework for mining data on Grid plaform [15, 9, 22] only propose services to handle distributed resources, process and knowledge. None of them offers any specific *DDM* algorithm. Besides, the integrating of traditional *DM* algorithms in these framework raises a question about the performance in real-world applications.
- Offering an effective knowledge management. As mentioned above, the process of local mining (build local knowledge) and then integrating all the results to create new knowledge is seen to be one of the most effective solutions for mining applications distributed over Grid platforms. This will lead to the problem of managing efficiently the mined results, so called knowledge, which becomes more and more complex and sophisticated. This is even more critical when the local knowledge of different sites are owned by different organisations. Besides, two steps of *DDM* (local mining and integrating) are not independent since naive approaches to local analysis may produce incorrect and ambiguous global data models. In order to take advantage of the mined knowledge at different locations, *DDM* framework should have a view of the knowledge that not only facilitates their integration but also minimises the effect of the local results on the global models. Briefly, an efficient management of distributed knowledge is one of the key factors affecting the outputs of these techniques.
- Managing efficiently distributed resources across Grid environment. Actually, most of the Grid Data Mining Projects in literature are based on Globus ToolKit (GT) [15, 9]. The use of a set of Grid services provided by this middleware helps the developer to deal with heterogeneous of resources. However, they also depend on GT's performance and problems such as security overhead. An efficient *DDM* framework needs to be more flexible with regards to many existing data grid platform developed today.
- Giving an user-friendly interface that helps users to build their mining applications and evaluate mining results easily and transparently over Grid platform.

In this section, we have just discussed important issues related to *DDM* systems. In the following sections, we will present a new *DDM* system for developing novel and innovative data mining techniques to deal with very large and distributed heterogeneous datasets in Grid environment.

## 3 New System Architecture

This architecture includes three main layers: core, virtual data grid and interface. In this section we present the first two layers. Meanwhile, the interface layer will be described in the Section 6.

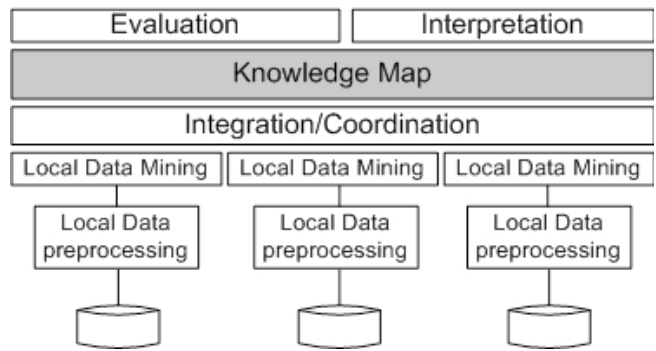

**Fig. 1.** System's architecture

### 3.1 Core layer

The core layer is composed of three components: knowledge discovery, task management and data/resource management.

The role of the knowledge discovery component is to mine the data; integrate and consolidate the data; and discover new knowledge. It is key component of this layer and it contains three modules: data preprocessing; distributed data mining (*DDM*) with two sub modules: local data mining (*LDM*) and integration/coordination; knowledge map.

The first module carries out locally data pre-processing of a given tasks such as data cleaning, data transformation, data reduction, data project, data standardisation, data density analysis, etc. These pre-processed data will be the input of the *DDM* module. Its *LDM* component performs locally data mining tasks. The specific characteristic of our new system compared with other current *DDM* systems is the ability of integrating different mining algorithms in a local *DM* task to deal with different kind of data. The local results will be integrated and/or coordinated by the second component of DDM module to produce global models. Distributed algorithms of for mining data are heart of the system and some of them will be presented in the section 4 below. At the interface level, user can choose *DDM* algorithms from a set of pre-defined ones in the system. Moreover, users can publish new algorithms to increase the performance.

The results of local DM such as association rules, classification, and clustering, etc. should be collected and analysed by domain knowledge. This is the role of the last module: knowledge map. This module will generate significant, interpretable rules, models and knowledge. Moreover, the knowledge map also controls all the data mining process by proposing different strategies for mining as well as for integrating and coordinating all the jobs to achieve the best performance. Details of the knowledge map can be found in the section 5.

The task management component plays an important role in this system. It manages all the schedules created from the interface layer. This reads an executing schema from the task repository and then schedules and monitors the execution of corresponding tasks. According to the scheduling, this task management component carries out the resource allocation and then finds the best and appropriate mapping between resources and task requirements. This part is based on services supplied by Data Grid layer (e.g. DGET [40]) in order to find the best mapping. Next, it will activate these tasks (local or distribution). This component is also responsible for the coordination of the distributed execution that is, it manages communication as well as synchronisation between tasks in the case of cooperation during the preprocessing, mining or integrating stages.

The role of data/resource management component is to facilitate the entire *DDM* process by providing an efficient control over remote resources in a distributed environment. This component creates, manages and updates information about resources in the dataset repository and the resource repository. The data/resource management component goes with the data grid layer to provide an transparent access to resources across heterogeneous platforms.

## 3.2 Data Grid layer

The upper part of this layer, called virtual data grid, is a portable layer for data grid environments. Actually, most of the Grid Data Mining Projects in literature are based on Globus ToolKit (GT) [15][9]. The use of a set of Grid services provided by this middleware gives some benefits. For instance, the developer do not waste time for dealing with heterogeneity of organisations, platforms, data sources, etc.; distributing of software is more easier because GT is the most widely used middleware in Grid community. However, this approach depends on GT problems such as security overhead, GT's organisation of system topology.

In order to make our system more portable, and more flexible with regards to many existing data grid platforms developed to date, we build this portable layer as an abstraction of virtual Grid platform. It supplies a general services operations interface to upper layers. It unifies different grid middle-wares by mapping *DM* tasks from upper layer to grid services according to OGSA/WSRF standard or to entities [40] in DGET model. The portable layer implements two groups of entities: data and resource entities. The first group deals with data and meta-data used by upper layers and the second deals with resources used. The DGET system guarantees the transparent access of data and resources across any heterogeneous platform. By using this portable layer, our system can be carried easily on many kind of Data Grid platforms such as GT and DGET.

## 4 Distributed Algorithms for mining large datasets

The system should not only be a platform based on *Grid*[32] infrastructure for implementing *DDM* techniques but also it should provide new distributed algorithms for exploring very large and distributed datasets. The first step of the development of these algorithms concern distributed clustering techniques which are well studied by the community in comparison with distributed association rules and distributed classification. In this section, we present some important *DDM* algorithms of clustering, frequent items set generation that can be integrated in the system. Other traditional *DM* algorithms such as K-means clustering and a range of its variants are also implemented in our system.

### 4.1 Variance-based Clustering

Clustering is one of the basic tasks in the data mining area. Basically, cluster-ing groups data objects based on information found in the data that describes the objects and their relationships. The goal is to optimise similarity within a cluster and the dissimilarities between clusters in order to identify inter-esting structures in the underlying data. There is already a large amount of literature on clustering ranging from models, algorithms, validity and perfor-mances studies, etc. However, there are still several open questions about the clustering process. These include:

- What is the optimal number of clusters?
- How to assess the validity of a given clustering strategy?
- How to allow different shapes of the clusters rather than spherical shapes generated by the given distance functions?
- How to prevent the algorithms initialization and the order in which the features vectors are read from affecting the clustering output?
- How to find which clustering structure for a given dataset, i.e why would a user choose an algorithm instead of another?

Answering these questions appropriately will guarantee the success of a clustering algorithm. Several algorithms have been developed to find several kinds of clusters (spherical, linear, dense, drawnout, etc.) depending on the data and its application.

In distributed environments, clustering algorithms have to deal with additional issues of distributed datasets, large number of nodes and domains, plural ownership and users, and scalability. It has been stated before that moving the entire data to a single location for performing a global clustering is not always possible due to different reasons. Moreover, communication issues are the key factors in the implementation of any distributed algorithm. It is obvious that a suitable algorithm for high speed network can be of little use in WAN-based platforms. Generally, it is considered that an efficient distributed algorithm minimises the data exchange and tries to avoid synchronisations as much as possible.

For this purpose, lightweight distributed clustering techniques are the best choice for these systems. This was shown to improve the overall clustering quality and finds the number of clusters and the global inherent clustering structure of the global datasets. The variance-based clustering is developed in this way and it is presented below.

## Algorithm foundations

The most used criterion to quantify the homogeneity inside a cluster is the variance criterion, or sum-of-squared-error. The traditional constraint used to minimize this criterion is to fix the number of clusters to an apriori known number, as in the widely used k-means and its variants [71], [56], [73], etc. This constraint is very restrictive since this number is most likely not known in most cases. However, many approximation techniques exist including the gap statistic which compares the change within cluster dispersion with that expected under an appropriate reference null distribution [68], [55], or the index of Calinski & Harabasz [14], among many others. The imposed constraint here states that the increasing variance of the merging, or union, of two subclusters is below a given dynamic limit. This parameter depends on the dataset and is computed using a global assessment method. This allows to find the proper value of the variance increasing by varying it without violating the locality principle of this algorithm. This parameter can also be available from the problem domain for a given data.

The key idea behind this algorithm is to choose a relatively a high number of clusters in local sites which are referred to as subclusters. An optimal local number of clusters using approximation techniques can be considered. Then, the global merging is done according to an increasing variance criterion requiring a very limited communication overhead. The algorithm finds the proper variance criterion for each dataset based on a statistical global assessment. This preserves the locality criterion for each dataset.

In each node (site), the clustering can be done using different algorithms depending on the characteristics of the dataset. This may include k-means, k-harmonic-means, k-medoids, the statistical interpretation using the expectation-maximisation algorithm, etc. The merging of local subclusters exploits the locality in the feature space, i.e. the most promising candidates to form a global cluster are subclusters that are the closest to each other in the features space. Each node can perform the merging and deduce global clusters, i.e. which subclusters are subject to form together a global cluster.

Another notion used in this algorithm is the border of a global cluster which represents local subclusters at its border. These subclusters are susceptible to be isolated and added to another global cluster in order to contribute to an improvement of the clustering output with respect to the variance criterion, i.e. that minimises the sum-of-squared-error. These subclusters are referred to as perturbation candidates. The initial merging order may affect the clustering output, as well as the presence of non well-separated global clusters. This process is intended to reduce the input order impact. The global clusters are then updated. The border is collected by computing the common Euclidean distance measure. The $b$ farthest subclusters are then the perturbation candidates, where $b$ is deduced depending on the local number of subclusters at each site and their global composition. Multi-assigned subclusters are naturally affected by this process.

The aggregation part of the algorithm starts with $\sum_{i \in s} k_i$ subclusters, where $s$ is the number of nodes involved and $k_i$, for $i = 1, ..., s$, are the local numbers of clusters in each node. Each node has the possibility to generate a global merging. An important point here is that the merging is a labeling process, i.e. each local node can generate the correspondences between local subclusters, without necessarily constructing the overall clustering output. This is because the only bookkeeping needed from the other nodes are centers, sizes and variances of the clusters. The aggregation is then defined as a labeling process between local subclusters. No data move is needed at this stage. On the other hand, the perturbation process is activated if the merging operation is no longer applied. The perturbation candidates are collected for each global cluster from its border, which is proportional to the overall size composition as quoted before. Then, this process moves these candidates by trying the closest ones first and with respect to the gain in the variance criterion when moving them from the neighboring global clusters. Formal definitions of this algorithm, and all notions and criterions, are given in [4].

## Complexity and evaluation

The complexity of this distributed algorithm depends on the algorithm used locally, the global assessment algorithm, the communication time which is a gather operation, and the merging computing time. If the local clustering algorithm is k-means, the clustering complexity is $O(N_{max}k_{max}d)$, where $d$ is the dimension of the dataset, i.e. number of attributes. The assessment complexity depends on the size of local statistics. If the gap statistic is used on local centers, this will be $O(B(\sum_{i \in s} k_i)^2)$, where $B$ is the number of reference distributions. The communication time is the reduction of $3d \sum_{i \in s} k_i$ elements. If $t^i_{comm}$ is the communication cost of moving one element from node $i$ to the aggregation node $j$, then the communication complexity is $3d \sum_{i \in s, i \neq j} t^i_{comm} k_i$. Since $k_i$ is much smaller than $N_i$ ($k_i \ll N_i$), the generated communication overhead is very small.

The merging process is executed a number of times, say $u$. This is the number of iterations until the merging condition is no longer applied. This cost is then equal to $u \times t_{newStatistcs} = O(d)$. This is followed by a perturbation that costs $O(bk_g k_{max})$. This process computes for each of the $b$ chosen subclusters at the border of a given cluster $C_i$, $k_g$ distances for each of the $k_g$ global clusters. The total cost is then $O(dN_i(\sum_{i \in s} k_i)^2)$ (with $T_{comm} \ll O(N_i k_i d)$).

This algorithm was tested on a range of simulated and real datasets including large Gaussian distributions, the well-known Iris dataset, the animal dataset, and the PUMS census dataset available from the UC Irvine KDD Archives. The algorithm finds the right global number of clusters by varying the maximum variance constraints, independently of the local clustering algorithm and the number of subclusters. An example using the Iris dataset, where the maximum variance constraint was twice the highest individual variance, is shown in Figure 2. In this case, since the k-harmonic-means does not impose a variance constraint it finds a lower sum-of-squared-error locally. However, the variance-based clustering finds the 3 initial classes based on 5 and 7 subclusters locally. More experiments and evaluation are shown in [4] and [6].

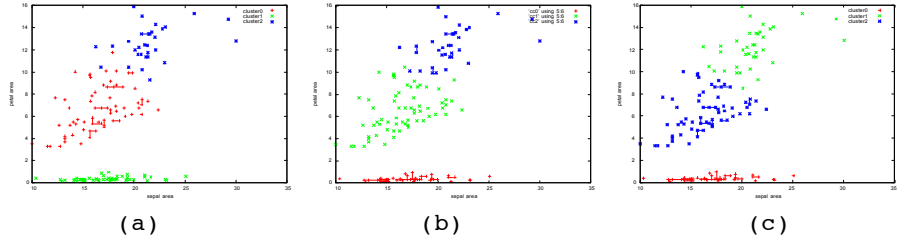

**Fig. 2.** The output using 5 (a) and 7 (b) subclusters, and a centralised clustering using k-harmonicmeans in (c).

### 4.2 Distributed Density-Based Clustering (*DDBC*)

Density-based clustering approaches have been widely used in mining large datasets. Moreover, density based clustering algorithms have been recognized to be powerful and capable of discovering arbitrary shapes of clusters as well as dealing with noise and outliers. In the developing of new distributed algorithms for integrating in the *DDM* system, a clustering approach based on density is also proposed. In this approach, the aggregating process is based on a decentralized model and the local clustering is a density-based. There are some density based algorithms in the literature such as DenClue[39] and DBSCAN[30]. In this approach, DBSCAN is chosen because it is simple and efficient in very large databases. It requires a minimum domain knowledge to determine input parameters and discover clusters with arbitrary shapes[30].

## Related researches on $DDBC$

In spite of a large amount of research conducted in distributed clustering such as [45][72][65], there are very few algorithms proposed in distributed density based clustering. Until now, to the best of our knowledge, there are four approaches in this paradigm that were presented in[70][42][43] and [50]. The former deals with a parallel approach of DBSCAN algorithm. This approach is appropriate for shared memory or distributed shared memory systems. The last three approaches include two main steps: local clustering to create local model and processing these local models to rebuild a global model.

In [42], authors used DBSCAN as a local clustering algorithm. They extended primitive elements of this algorithms such as core points, $E$, $Minpts$ [30] by adding new concepts as specific core points, specific $E_{range}$ to build a local representative at each site. The global model will be rebuilt by executing the DBSCAN algorithm on a set of local representatives with two global values: $M intps_{global}$ and $E_{global}$. $M intps_{global}$ is a function of two local parameters i.e. $M intps_{global}$= 2 x $Minpts$. $E_{global}$ is tunable by the user and its default value is the maximum value of all $E_{range}$ values of all local representatives. This approach has some advantages: firstly, local clustering can be executed rapidly and independently at each site. Secondly, by using local representatives, it decreases the communication cost by avoiding to send all datasets and therefore the global clustering can be done quickly. However, this approach has two crucial problems: it ignores the local noise and the default value is set to $E_{global}$. There is no representation of noise in the local representatives. In the global view, local noise from one site can belong to one or many clusters of other sites. Moreover, a set of local noises from all local sites can form one or more new clusters. Furthermore, the use of the high and static value $E_{global}$ led to incorrect cluster merging as shown in [42]. In addition, the location of special core point may also effect the merging process when they are located at the border of cluster. Another approach proposed in [43] has also two main steps as in the first approach but the definition of local representatives is based on the maximum distance between a representative and its covered objects. This approach has same advantages as the first one. Moreover, it can tackle not only with the problem of noise but also border problem as mentioned above. However, choosing a suitable number of local representatives is difficult task.

The new approach presented in this section is also composed of two important steps: local clustering to create local models and hierarchical agglomeration of local models to build a global model. These two steps use different algorithms as in [42][43]. We have also the pre-processing and post-processing stage. For the convenient, we define firstly the convention of symbols used in the next sections: (i) the letters *x, y, z* are reserved for local sites e.g. site x, site y; (ii) *i, j, k, l*: index of elements in a set; (iii) *t, u, v*: number of elements in a set; (iv) *s, c, n*: elements; (v) *S, C, CorC, N, A, L*: sets; (vi) *E, δ, θ*: value or threshold.

## Local clustering

All sites carry out a clustering process independently from each other to discover local models. The local clustering algorithm chosen in our approach is DBSCAN, because it is strong approach concerning outliers and it can be used for all kinds of metric data space and vector spaces and it is simpler than other density-based algorithms e.g. DenClue[39]. At each site, the local model created consists of a set of representatives. Choosing a set of representative is very important because it will affect the performance of the merging step as well as the accuracy of the global model built. This depends normally on local mining algorithm. In DBSCAN algorithm, core points w.r.t $E$ and *Minpts* play an important role in the determination of different clusters. We can naturally use core points w.r.t $E$ and *Minpts* as representatives. However, the number of core points is not small enough with regard to the number of data points. Using core points is not efficient in the case of large amount of local data. We will use, instead a set of absolute core points $S_{cor}$ w.r.t $E$ and *Minpts* as the first part of our representatives. Let $CorC_i^x \subseteq C_i^x$ (cluster $C_i^x$ in a set of cluster $C^x$ at site x) be a set of the core points belonging to this cluster at site x. The definition of $S_{cor}$ is as follows: $S_{corC_i^x} \subseteq CorC_i^x$ is a set of absolute core object iff ($\forall s_k, s_l \in S_{corC_i^x} : s_k /= s_l \Rightarrow s_k /\in N_{E_x}(s_l)$) and ($\forall c \in CorC_i^x, \exists s \in S_{corC_i^x} : c \in N_{E_x}(s)$) with $N_{E_x}(p_i)$ of a point $p_i$ is defined as $\forall p \in N_{E_x}(p_i) : p_i - p' \leq E_x$.

The distance used is either Euclidian or Manhattan distance. The concept of absolute core point was also proposed in [42] where it is called special core point. In the definition of an absolute core point $s$, there is at least one core point within the range of $s$ w.r.t $E$ and *Minpts*. So we also add the furthest core point within the range of $s$ w.r.t $E$ and *Minpts* in the first part of our local representative. Briefly, this first part is $R^x$ which includes clusters representative $R_i^x$ containing their set of pair (absolute core point $s$, its farthest touchable core point $c_s$):

$$R^x = \{ \underset{i:1..t}{R_i^x} \mid R_i^x = \underset{j:1..u}{\{(s, c_s) \mid s \in S_{corC_i^x}\}}\} \tag{1}$$

The Fig. 3a shows an example of absolute core point. This approach is different from [42] using only an $E$-range value for each absolute core point. One of the reasons of using this value is to deal with data points in the range of a core point (w.r.t $E$ and *Minpts*) and this core point is the furthest of an absolute core point (Fig. 3b). However, using this value might lead into the problem of merging clusters that are not similar. It can happen when the absolute core is on the border of its cluster (Fig. 3c). In order to tackle this problem, we do not use this value but adding the furthest core point of an absolute core point in our local representative. The second part of it is the value $E_x$ which is the local $E_x$ value and it can be estimated as discussed in [30]. These local values are different between local sites. Finally, the last part is a set of noise data $N^x$ in this local site. The noise at one local site might belong to a cluster at other sites and moreover the aggregating of noise from all local sites might create one or few new clusters (Fig. 3d). The noise at a local site is defined as data points that is not belong to any cluster in this local site.

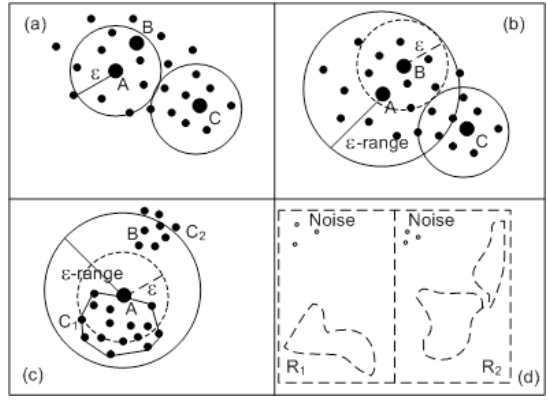

**Fig. 3.** (a) A & C are Absolute Core points;(b) B is the furthest core point of A;(c) A is on the border of cluster $C_1$: by using $E$-range, two clusters $C_1$, $C_2$ will be merged ;(d) merging of noise from two local sites R1, R2 might create a new cluster

$$N^x = \{ \underset{j:1..v}{} d_j \mid d_j \, / \in C_i^x : \forall C_i^x \in C^x \} \qquad (2)$$

Then, our representative of local model is defined as follows:

$$LocalM\, odelL^x = \{R^x, E_x, N^x\} \qquad (3)$$

In the next sub-section, we present the merging process of local clusters to obtain global model.

**Hierarchical-agglomeration of local model to build global model based on tree topology**

The process of merging local models is based on tree structure (binary or Tree-P[28] Fig. 4b). At lowest level (leaf-level), each node of a tree is a physical site where a local model is stored. Nodes at higher levels are virtual sites. At the beginning, local models of a local site are stored at a leave node. These local models are merged by binary tree or by group (Tree-P) into sub-global model at its logical site. For instance, as in Fig. 4a, local models from site X and Y will be merged into a sub-global model at site z. These sub-global models are at their turn merged into other logical node at higher levels until we reach the root. At root, we have a global model. The group merging is essentially based on merging two or more local models. By using tree topology, we can implement not only our approach on Data Grid platforms such as DGET[40] but also avoid the problem of bottle-neck in traditional client-server approach. Moreover, in tree-based topology, we can stop the merging process at any level.

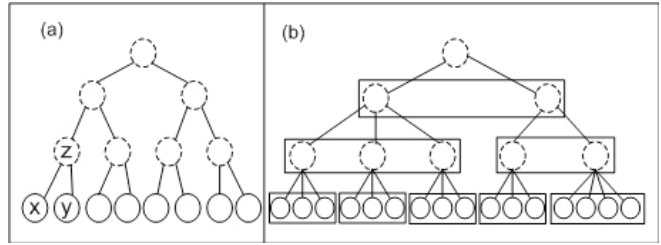

**Fig. 4.** (a) Binary Tree and (b) Tree-P topology

This algorithm assumes that the local clustering at all of the sites uses the same value of *Minpts*. Meanwhile, each site $x$ has its own value of epsilon $E_x$. Normally, the merging process needs a global epsilon. However, finding a suitable global epsilon value $E_{global}$ is a difficult task. The higher $E_{global}$ is the more risk of merging wrong clusters. In [42], authors proposed a tunable $E_{global}$ value depending on $E$-range values of all local representatives. An $E$-range value is composed of $E$ value and a distance between an absolute core point at its furthest core point. And all of the local sites use the same $E$ value. In the Grid environment where datasets are produced and processed by a large number of different owners, suppose that they use the same mining algorithm, e.g. DBSCAN it is difficult to have the same parameter e.g. $E$ for all sites. Actually, this new algorithm uses different value of epsilon $E_x$ for each local site $x$ and the global $E_{aver}$. This $E_{aver}$ is determined as shown in [49].

***Merging Process***

Suppose that at site $z$, we will merge local models $L^x$ and $L^y$ from two sites $x$ and $y$ to build a new sub-global model $L^z$. If z is a root site, this sub-global model becomes the global model that should be returned. The term "cluster" used in this section means a set of representatives of that cluster.

Firstly, we find the minimum value of epsilon $E$ from two sites as $E = \min(E_x, E_y)$. We have also to solve with the problem of the difference $\delta$ of epsilon value between two sites: $\delta = |E_x - E_y|$. If $\delta$ is small enough, we could merge directly two clusters, this case is called a direct aggregating. If $\delta$ is too large, we can only merge separately representatives of the site for which the epsilon value is the largest. Some of their representatives might be disaggregated and then will be merged with other clusters. This case is called a disaggregating of cluster.

We define a threshold of disaggregating $\theta$. If $\delta$ is less than $\theta$ then we are in the first case, else we are in the second case. Moreover, an $E_{aver}$ is be used instead of $E$ in the second case. We propose a simple method to determine both $\theta$ and $E_{aver}$ as shown at the end of this section.

We assume that the local model at site $x$ contains the minimum $E$ value without loss of generality. The merging process has two important steps: cluster extension and cluster merging. In the first step, each cluster from site $x$ will be extended by adding noise data from $N^y$. A noise $d_i \in N^y$ will be included in the cluster $R_i^x \in R^x$ if:

$$\exists (s, c_s) \in R_i^x : \quad d_i - s \quad \leq E \lor \quad d_i - c_s \quad \leq E \tag{4}$$

Let $A_i^x$ be a set of noise data from $N^y$ that belongs to the cluster $R_i^x$. After the first step we have a set $A_i^x$. We execute the same process with each cluster from site $y$ by adding noise data from $N^x$ if we are in the direct aggregating case and we will also have a set $A_i^y$.

In the second step, there are two cases that correspond to aggregating-disaggregating cases defined above. In the direct aggregating case, we will merge two clusters $R_i^x$ and $R_j^y$, if $\forall R_i^x \subseteq R^x$, $\forall R_j^y \subseteq R^y$, $\exists (s_k, c_{sk}) \in R_i^x$, $\exists (s_l, c_{sl}) \in R_j^y$:

$$s_k - s_l \quad \leq E \lor \quad c_{sk} - s_l \quad \leq E \lor \quad s_k - c_{sl} \quad \leq E \lor \quad c_{sk} - c_{sl} \quad \leq E \tag{5}$$

The result of this merging will create a new cluster $R_k^z \subseteq R^z$. Meanwhile, in the disaggregating case, we only merge separately each representative $(s_l, c_{sl})$ of a cluster $R_j^y$ with cluster $R_i^x$ if it satisfies the equation (5) by using of $E_{aver}$ instead of $E$.

These representatives will be removed from $R_j^y$ and included in the set $A_i^x$. After two step of merging the new local model $L^z$ is created:

$$L^z = \{R^z, E_z, N^z\} \tag{6}$$

with:

- $E_z = E$ or $E_{aver}$
- $R^z = \quad _{\forall (i,j)} (R_i^x \quad R_j^y) \quad _{\forall i} A_i^x \quad _{\forall j} A_j^y$, if $R_i^x$ and $R_i^y$ satisfies equation (5) and $N^z = N^x \quad N^y$ for aggregating case or
- $R^z = \quad _{\forall i} (R_i^x \quad A_i^x) \quad (\forall (s_l, c_{sl}) \in R_j^y) : (s_l, c_{sl}), R_i^x$ satisfies equation (5) and $N^z = N^x \quad N^y$ for disaggregating case.

This model will be used to continue the merging process with the local model from another site depending on the topology chosen until we obtain a global model. Moreover, as we mentioned above, this process can stop at any level of the tree topology and the sub-global models are returned as the final results.

## Complexity and Evaluation

Details about this algorithm can be founded in [49]. In this sub section, we briefly present the complexity of this algorithm and its preliminary experiments.

We suppose that there is a total of N data points divided equally among m computing nodes. So, each node has n = N/m data points. We also assume that the number of representatives in one node is approximately $\mu$% of the total data points of this node. The complexity of our approach is composed of two parts: local mining and global aggregating. The local mining is based on DBSCAN algorithm so its complexity is $O(n \log n)$[30]. The complexity of global aggregating is $(n\mu)^2 \log m)$. Briefly, the complexity of our approach is $O(n \log n + (n\mu)^2 \log m))$. The speedup compared with centralization approach is:

$$S_p = \frac{(m-1)\log N}{(\mu^2 m - 1)\log m}, with(\mu^2 m - 1) \geq 0 \qquad (7)$$

This speed up depends on $\mu$ and number $m$ of computing nodes. The more nodes we use and the less number of representatives that we can have, the more speed up that we could gain for the same dataset.

Experiments of this algorithm are launched with datasets from LOCAL project [52]. This datasets includes 322 objects of two dimensions. Firstly DBSCAN algorithm is executed on this datasets with $E$=0.004 and $Minpts$=4 and there are six clusters as shown in Fig. 5. Next, this datasets are distributed equally by round robin in two subsets. Then, DBSCAN algorithm is separately executed on each subset with $E$=0.005 and $Minpts$=4. The local $E$ value is chosen based on the local distribution of each subset. As shown in Fig. 6, there are 7 clusters for the first subset (Fig. 6 left) and 8 clusters for the second (Fig. 6 right).

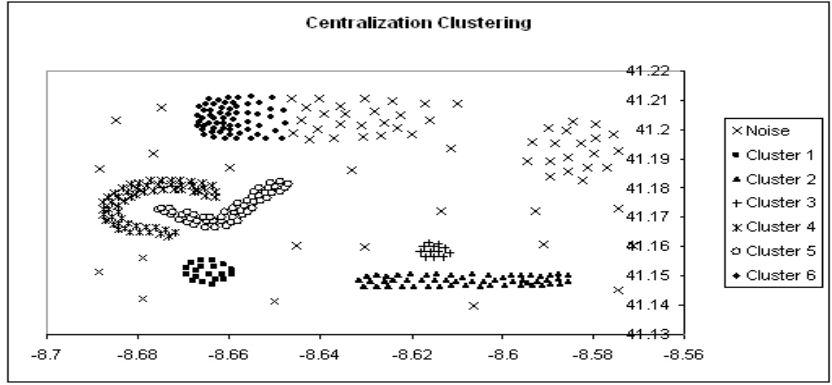

**Fig. 5.** Centralization Clustering

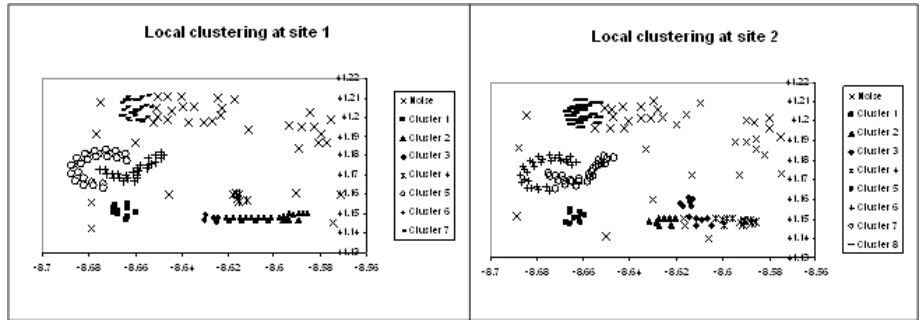

**Fig. 6.** Local clustering at two sites

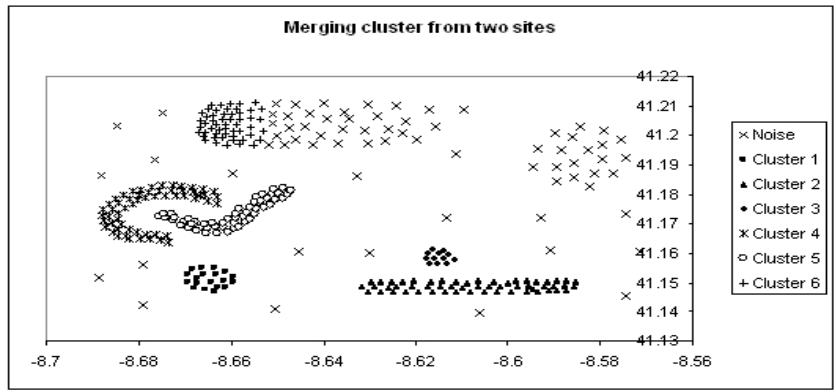

**Fig. 7.** Global merging

The next step is to build the local representatives for each subset of data. Then, the merging process is executed and merging result is shown as in Fig. 7. Note that we obtained the same number of clusters as in centralized clustering but some data points of cluster 6 became noise points. The reason is that these points were apparently noise points in local clustering. Moreover, the quality of distributed clustering (by using the ***continuous object quality P*** as proposed in [42]-page 100) obtained from these experiments is around 94.43%. The result shows that the merging process is efficient.

### 4.3 Distributed Frequency Itemsets Generation

The frequent itemset mining task is at the core of various data mining applications. Since its inception, many frequent itemset mining algorithms have been proposed [2], [11], [38], [61], [64], among many others. Many of them are based on the Apriori or the FP-Growth principals. Basically, frequent itemsets generation algorithms analyse the dataset to determine which combination of items occurs together frequently. For instance, considering the commonly known market basket analysis; each customer buys a set of items representing his/her basket. The input of the algorithm is a list of transactions giving the sets of items among all existing items in each basket. For a fixed support threshold $s$, the algorithm determines which sets of items of a given size $k$ are contained in at least $s$ transactions.

The focus here is on mining frequent itemsets on distributed datasets over the grid. The grid-based approaches are motivated by the inherent distributed nature of these applications, and by the challenge of developing scalable solutions for the data mining field, which is highly computationally expensive and data intensive. Effective distributed approaches for large scale data mining should take into account both the challenges raised by the underlying grid system and the complexity of the task itself. For the purpose of developing well-adapted grid implementations, a performance study of frequent itemsets mining of large distributed datasets on the grid, based on the Apriori principle is required.

The study of the distributed aspect and the performance of Apriori-based approaches is done both theoretically and experimentally. The theoretical study presents a performance model of distributed algorithms based on the Apriori principal. Note that the main factor of an Apriori-based distributed algorithm is the number of candidates generated at each step or level. This factor, which governs the algorithm complexity, can be exponential of the size of the input. In the distributed version of this type of algorithms, one tries to maximise the number of concurrent activities (or parallel activities) and reduce the overheads of the communications and synchronisations. We show that local pruning strategies are sufficient and that global phases in classical distributions affect directly the performance of the system.

The approach introduces in [7] has two main phases. The first phase consists of generating frequent itemsets on each node based only on their local datasets. This phase is a *local mining phase* and it uses the traditional sequential Apriori algorithm. After this phase, the result will be the set of all locally frequent itemsets in each node. This information is sufficient for determining all globally frequent itemsets, using a top-down search. The second phase is the *global collection phase*. Each node broadcasts its frequent itemsets, of size $k$ and maximal ones, to the others nodes of the system and asks for their respective support counts. The globally frequent itemsets are then identified by merging local support counts from each node. Then, the algorithm iterates on the subsets of itemsets that fail the global frequency test. More precisely the globally frequent itemsets are generated as follows:

1. Initially collect support counts of frequent itemsets of size $k$ (the requested size) and all smaller frequent itemsets that are not subsets of any larger frequent itemset (maximal itemsets).

2. Generate globally frequent itemsets, and put all the itemsets that are not globally frequent in a set $F$.
3. If $F$ is not empty, collect support counts of subsets of itemsets in $F$ and go to (2).

This top-down search has been shown to be efficient, and the overheads due to synchronisations and communications are significantly reduced. This leads to much fewer communication passes. Also, the global pruning steps add extra computational costs in local nodes and therefore affects the global system performance.

## Discussion and evaluation

Comparisons with a classical Apriori-based distributed approach, namely the Fast Distributed Mining of association rules (FDM), show that in terms of computation, both algorithms perform approximately the same amount of work as they have the same amount of candidates in the local Apriori generation. However, in terms of communication, the proposed top-down approach performs better and has only two communication passes on both synthetic and real datasets; the PUMS census dataset, and datasets generated using the IBM Quest code respectively. The IBM Quest code is a simulation model for supermarket basket data. It has been used in several frequent itemsets generation studies such as [38], [60], and [66], etc.

As example, the Figure 8 shows plots of different candidates sets on different nodes using various support thresholds, on the two mentioned datasets. The lower bound, which is the ratio between the number of candidate sets of the two techniques, is 0.78. This value is close to 1 in most cases, with an average value of 0.93. If we look at the ratio of the number of 1-itemsets for the two techniques we can see the same behaviour with an average value of 0.94. One can conclude that the difference in terms of candidate set generation between the two techniques is not significant. In terms of processing time, this is in the order of few seconds in all cases. For the overall computation costs, the proposed technique has a gain factor of up to 82%. However, this highly depends on the size of frequent itemsets and the number of communication passes. Also, the input and output requirements were not considered in this model for simplicity. This is likely to be more costly in the case of classical distributed approaches since the proposed approach generates less important overall sets for remote support count collection.

Basically, the results show that distributed implementations of the Apriori algorithm do not need global pruning strategies. Therefore, classical distributions are less efficient than the adopted global strategy in our approach, starting from the requested size and using a top-down search. Note that remote support counts computations can be very expensive in classical distribution, especially in lower levels where the number of locally frequent itemsets is high. This was avoided and reduced to a minimum in the proposed approach since only a few passes of remote computations are required and with smaller sizes. Formal definitions and more detailed results are presented in [7]. This method is intended not only to reduce synchronisation and communication overheads but also the grid tools overheads which are due to jobs preparation or scheduling for instance. Efficient grid implementations should avoid multiple communication and synchronisation steps as much as possible.

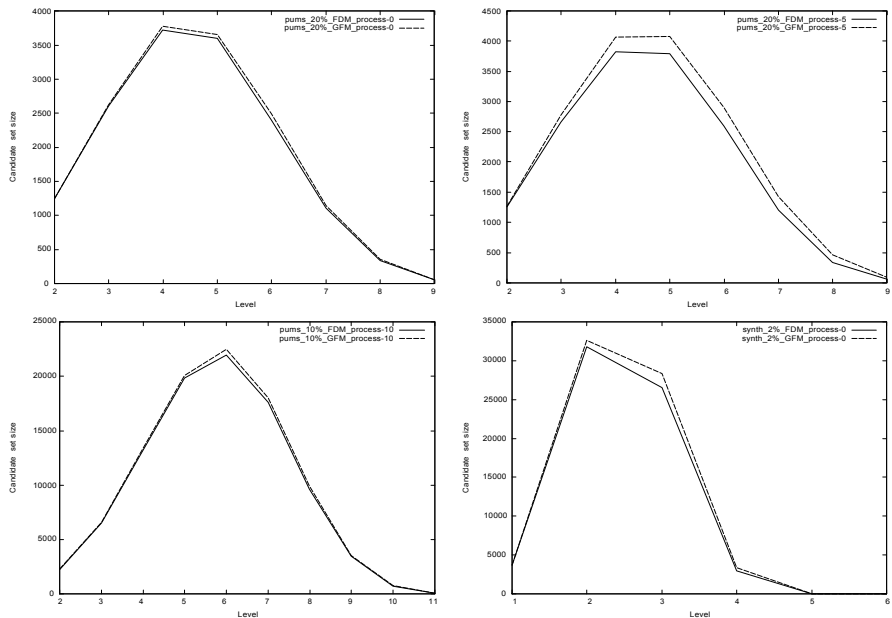

**Fig. 8.** The difference between some generated candidate sets using both approaches on different processes.

## 5 Knowledge Map

Today, the problem of managing efficiently the mined results, so called knowledge, which becomes more and more complex and sophisticated. This is even more critical when the local knowledge of different sites are owned by different organisations. Usually existing *DDM* techniques perform partial analysis on local data at individual sites and then generate global models by aggregating these local results. These two steps are not independent since naive approaches to local analysis may produce incorrect and ambiguous global data models. In order to take advantage of the mined knowledge at different locations, *DDM* should have a view of the knowledge that not only facilitates their integration but also minimises the effect of the local results on the global models. Briefly, an efficient management of distributed knowledge is one of the key factors affecting the outputs of these techniques.

A "Knowledge map" can be used to handle knowledge of *DDM* tasks on large scale distributed systems and also supporting the integration views of related knowledge. The concept of knowledge map has been efficiently exploited for managing and sharing knowledge [57] in different domains but not yet in *DDM* field. The main goal here is to provide a simple and efficient way to handle a large amount of knowledge built from *DDM* applications in Grid environments. This knowledge map helps to explore quickly any results needed with a high accuracy. This will also facilitate the merging and coordination of local results to generate global models. This knowledge map is also one of the key layers of the *DDM* system. This section starts with a background of knowledge representation and knowledge map concept. We present then architecture of knowledge map layer. Next, following paragraph deal with implementation issues. An evaluation of this approach is also presented to terminate this section.

## 5.1 From Knowledge Representation to Knowledge Maps

### Representation of knowledge mined

There are many different ways of representing mined knowledge, such as decision tables, decision trees, classification rules, association rules, instance-based, and clusters. Decision table [27] is one of the simplest ways of representing knowledge. The columns contain a set of attributes including the decisions and the rows represent the knowledge elements. This structure is simple but it can be sparse because of some unused attributes. Decision tree [27] approach is based on "divide-and-conquer" concept where each node tests a particular attribute and the classification is given at the leaves level. However, it has to deal with missing value problem. Classification rules' approach [27] is a popular alternative to decision trees. It uses production rules [12], called cause-effect relationships, to express the knowledge. Association rules [27] are kind of classification rules except that they can predict any attribute and this gives them the freedom to predict combinations of attributes too. Moreover, association rules are not intended to be used together as a set, like classification rules. The instance-based knowledge representation uses the instances to represent what is mined rather than inferring a rule set and store it instead. The problem is that they do not make explicit the structures of the knowledge. In the cluster approach, the knowledge can take the form of a diagram to show how the instances fall into clusters. There are many kinds of cluster representations such as space partitioning, Venn diagram, table, tree, etc. Clustering [27] is often followed by a stage in which a decision tree or rule set is inferred that allocates each instance to its cluster. Other knowledge representation approaches, such as Petri net [59], Fuzzy Petri nets [19] and G-net [26] were also developed and used.

**Knowledge Map Concept**

A knowledge map is generally a representation of "knowledge about knowledge" rather than of knowledge itself [25] [29] [69]. It basically helps to detect the sources of knowledge and their structures by representing the elements and structural links of the application domains. Some kind of knowledge map structures that can be found in the literature are: hierarchical/radial knowledge map, networked knowledge map, knowledge source map and knowledge flow map.

Hierarchical knowledge map, so-called concept map [57], provides a model for the hierarchical organization of the knowledge: top-level concepts are abstractions with few characteristics. Concepts of the levels below contain detailed traits of the super concept. The links between concepts can represent any type of relations as "is part of", "influences", "can determine", etc. A similar approach is radial knowledge map or mind map [13], which consists of concepts that are linked through propositions. However, it is radially organised (star topology). Networked knowledge map is also called causal map which is defined as a technique "for linking strategic thinking and acting, making sense of complex problems, and communicating with others what might be done about them" [13]. This approach is normally used for systematising knowledge about causes and effects. Knowledge source map [29] is a kind of organisational charts that does not describe functions, responsibility and hierarchy, but expertise. It helps experts in a specific knowledge domain. The knowledge flow map [29] represents the order in which knowledge resources should be used rather than a map of knowledge.

## 5.2 Knowledge Map Layer Structure

The knowledge map (*KM*) does not attempt to systematize the knowledge itself but rather to codify "knowledge about knowledge". In our context, it facilitates the deployment of *DDM* by supporting users coordination and interpretation of the results. The objectives of our *KM* architecture are: 1. provide an efficient way to handle a large amount of data collected and stored in large scale distributed system; 2. retrieve easily, quickly, and accurately the knowledge; and 3. support the integration process of the results. A *KM* architecture is proposed as shown in Fig.9, 10 and 11 to achieve these goals. *KM* consists of the following components: knowledge navigator, knowledge map core, knowledge retrieval, local knowledge map and knowledge map manager (Fig.9). From now on, we use the term "mined knowledge" to represent for knowledge built from applications.

**Knowledge navigator**

Usually, users may not exactly know what they are looking for. Thus, knowledge navigator component is responsible for guiding users to explore the *KM* and for determining the knowledge of interest. The result of this task is not the knowledge but its metadata, called ***meta-knowledge***, which includes related information such as data mining task used, data type, and a brief description of this knowledge and its location. For example, a user may want to retrieve some knowledge about tropical cyclone. The application domain "meteorology" is used by this component to navigate the user through tropical cyclone area and then a list of information related to it will be extracted. Next, based on this meta-knowledge and its application domain, the users will decide which knowledge and its location are to be retrieved. It will interact with knowledge retrieval component to collect all the results from chosen locations.

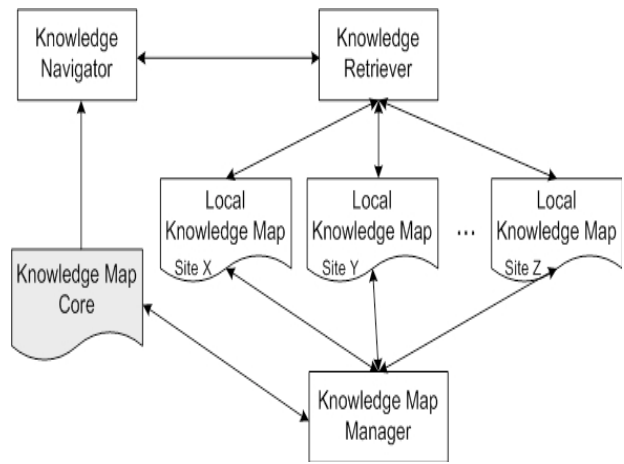

**Fig. 9.** Knowledge map system.

## Knowledge map core

This component (Fig.10) is composed of two main parts: *concept tree repository* and *meta-knowledge repository*. The former is a repository storing a set of application domains. Each application domain is represented by a *concept tree* that has a hierarchical structure such as a concept map [57]. A node of this tree, so called *concept node* represents a sub-application domain and it includes a unique identity, called *concept Id*, in the whole *concept tree* repository and the name of its sub-application domain. The content of each *concept tree* is defined by the administrator before using *KM*. The concept tree repository could also be updated during the runtime. In our approach, a mined knowledge is assigned to only one sub-application domain and this assignment is given by the user. As shown in Fig.10 for example, the concept tree repository contains an application domain named "meteorology" which includes sub-application domains such as "weather forecasting", "storm" and "climate". And then, "thunder storm", "tropical cyclone" and "tornado" are parts of "storm". By using *concept tree*, we can deal with the problem of knowledge context. For instance, given the distributed nature of the knowledge, some of them may have variations depending on the context in which it is presented locally.

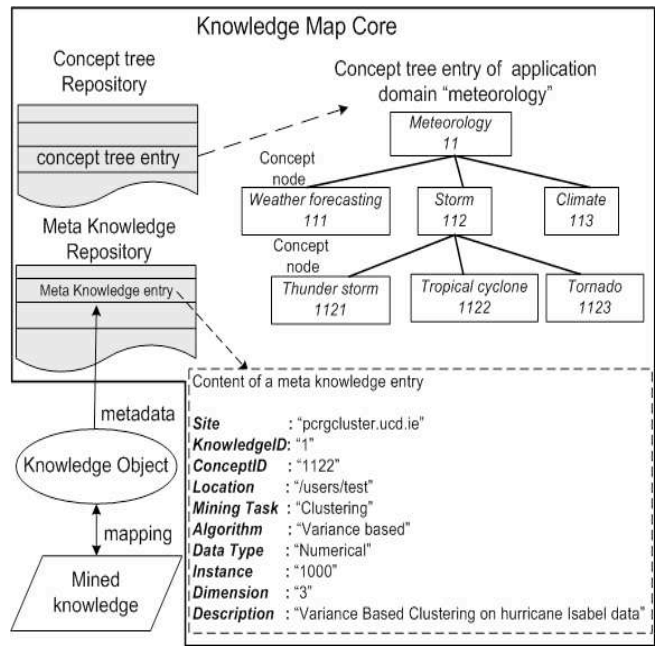

**Fig. 10.** Knowledge map core structure.

**Meta-Knowledge repository** (Fig.10): this handles metadata of the mined knowledge from different sites. A knowledge is mapped to a **knowledge object** and its metadata is represented by a meta-knowledge entry in this repository. Figure 10 also shows an example of a meta-knowledge entry. Based on this information, users could determine which mined knowledge they want to extract.

The goal of the *KM* core, is not only detecting the sources of knowledge and information but also representing their relationships with concepts of a given application domain. This component could be implemented in a master site depending on the topology of the system that will be discussed below.

**Knowledge retrieval**

The role of this component is to seek the knowledge that is potentially relevant. This task depends on the information provided by the users after navigating through application domains and getting the meta-knowledge needed. This component is similar to a search engine which interacts with each site and collects the local knowledge.

**Local knowledge map**

This component (Fig.11) is local to each site of the system. *Local knowledge map* is a repository of knowledge entries. Each entry, which is a knowledge object, represents a mined knowledge and contains two parts: *meta-knowledge* and a *representative*. *Meta-knowledge* includes information such as the identity of its mined knowledge that is unique in this site, its properties, and its description. Theses attributes were already introduced in the section ***Knowledge map core*** above. This *meta-knowledge* is also submitted to the ***Knowledge map core*** and will be used in *meta-knowledge entry* of its repository to be used at the global level. The *representative* of a knowledge entry depends on a given mining task. KM supports two kinds of representatives: one for clustering task and another for rule-based knowledge. Moreover, our system has the capacity of adding more representative types for other mining tasks.

For rule-based knowledge (Fig.11b), the mined knowledge is represented as a set of the production rules [12]. As mentioned above, a rule is of the form "IF {cause expression} THEN {conclusion expression}" and an expression (cause or conclusion) contains a set of items. A rule also includes its attributes such as *support* and *confidence* [27] in association rules task or *coverage* and *accuracy* [37] in classification task, etc. In order to represent these rules by their items, a representative in our approach consists of two parts: a *rule table* and an *item index table*. The former is a table of rules where each line represents a rule including its identity, content, attributes and creation information. The *item index table* is a data structure that maps items to the rule table. For example, the index of a book maps a set of selected terms to page numbers. There are many different types of index described in the literature. In our approach, the index table is based on *inverted list* [74] technique because it is one of the most efficient index structures [75]. This index table consists of two parts: items and a collection of lists, one list per item, recording the identity of the rule containing that item. For example in (Fig.11a), we assume that the term *"cloud"* exists in rules of which identities are *25, 171, 360*, so its list is *{25, 171, 360}*. This index table also expresses the relationship between items and their corresponding rules. By using this table, rules which are related to the given items will be retrieved by the intersection of their lists, e.g. the list of term *"pressure"* is *20, 171* so the identity (ID) of rule that contains *"cloud"* and *"pressure"* is *171*. This ID is then used to retrieve the rule and its attributes. In addition, a rule can be created by using one or more other rules, so its creation information keeps this link (see Fig.11c).

In the clustering case (Fig.11a), a representative of a mined knowledge stands in one or many clusters. A cluster has one or more representative elements and each element consists of fields filled by the user. The number of fields as well as data type of each field depends on the clustering algorithm used. The meta-data of these fields is also included in each representative. *KM* allows the user to define this meta-data. A cluster also contains information about its creation. This information shows how this cluster was created: by clustering or integration process. In the former case, the information is a tuple of (*hostname, cluster filename, cluster identity*) and in the latter, it is a tuple of (*hostname, knowledge identity, cluster identity*), where *hostname* is the location of the clustering results, which are stored as in files called cluster files with their *cluster filenames*. Each cluster has a *cluster identity* and it is unique in its knowledge entry. For example, a knowledge entry which is created by a variance-based clustering algorithm [5] on test datasets, has its representative as shown in Fig.12. In this example, there are four clusters, each one has only one representative. A cluster representative consists of three fields: *cluster identity*, *counts*, *centres* and *variances* with their data types which are *integer*, *long*, *vector 3 of doubles* and *matrix 3x3 of doubles* respectively. The content of a cluster representative is presented after its meta-data. Besides, another important information of cluster representative is the creation type which shows how this cluster was created: by either a clustering process or an integration process which merges sub-clusters from different sources. In the integration case, the cluster representative shows its integration link representing all information needed to build this cluster. Fig.11c shows an example of integration link. In this figure, the cluster at the root level is integrated from three other sub-clusters where the last one is also integrated from two others. Note that in Fig.11c, representatives (ii) and (iii) belong to the same knowledge.

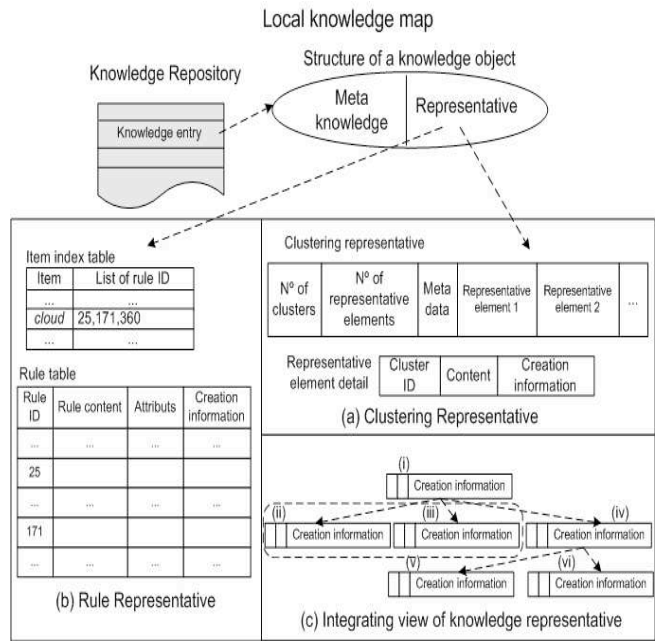

**Fig. 11.** Local knowledge map.

```
Representative type="Clustering"
   Number of clusters        : 3
   Metadata fieldnames:Id, Counts, Centres, Variances
   Metadata type  : int; long; double[3]; double[3][3]
   ==========================================
   ClusterID = 0
      ......
   ------------------------------------------------------------
   ClusterID = 1
      Id       : 1
      Counts   : 1500
      Centres  : 0.005,0.006,0.007
      Variances: 0.0051,0.0061,0.0071
                 0.0052,0.0062,0.0072
                 0.0053,0.0063,0.0073
      Creating type = "Integrating"
         Number of sub-elements: 3
         Sub-Element 1:...
         Sub-Element 2:...
         Sub-Element 3:
            site         : compute-0-1
            knowledgeID: 1
            clusterID    : 2
            Number of sub-elements: 2
            sub-element 3.1:
               site       : compute-0-0
               knowledgeID: 1
               clusterID   : 1
               number of sub-element: 0
            sub-element 3.2:
               site       : compute-0-2
               knowledgeID: 0
               clusterID   : 0
               number of sub-element: 0
   ------------------------------------------------------------
   ClusterID = 2
      ......
   ------------------------------------------------------------
```

**Fig. 12.**  A cluster  representative.

## Knowledge map  manager

Knowledge map manager is responsible for managing and coordinating the
local knowledge maps and the knowledge map core. For *local knowledge map*,
this component provides primitives to create, add, delete, update knowledge
entries and their related components (e.g. *rule net* and *item index table*) in
knowledge repository. It also allows to submit local meta knowledge to the its
repository in *knowledge map core*. This component provides also primitives
to handle the meta-knowledge in the repository as well as the concept node
in the concept tree repository. One key role of this component is to keep the
coherence between the *local knowledge map* and *the knowledge map core*.

### 5.3  Implementation of Knowledge Map

*KM* is implemented as a runtime including a set of *KM* daemons (Fig.13).
Each local site has one *KM*  daemon that is responsible for processing both
local and remote requests. One site of the system is chosen as a host to store
the *meta-knowledge repository*.

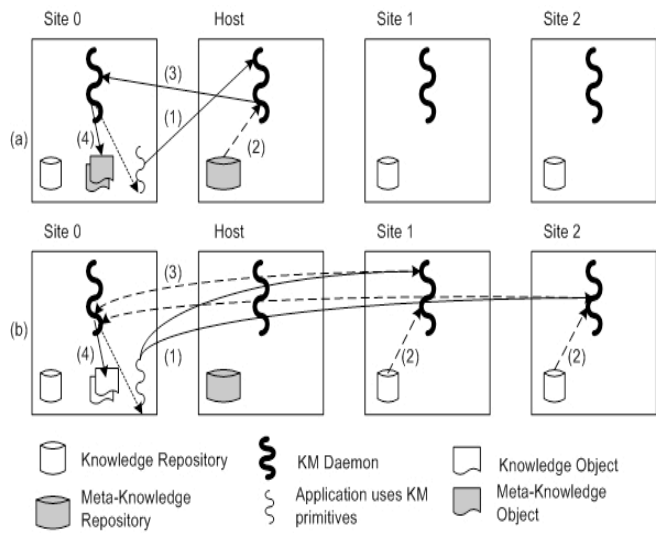

**Fig. 13.** An example of using Knowledge Map.

An example of searching/retrieving is shown in Fig.13. The search operation is composed of four steps: (1) a request is sent to the Host to look for the meta-knowledge needed. Then, this will be retrieved (2) and sent back to the source site (3), and it extracts the results as meta-knowledge objects (4). The retrieving operation is also composed of four steps: (1) requests are sent to the appropriate sites; (2) retrieve the knowledge found at each site; (3) send back to the source site via *KM Daemon*; (4) extracts results as knowledge objects.

### 5.4  Evaluation

This tool is used in the system described in [48][47]. It is difficult to evaluate our approach by comparing it to other systems because it is unique so far. Therefore, this new approach is validated by evaluating different aspects of the system architecture for supporting the management, mapping, representing and retrieving the knowledge. First, we evaluate the complexity of search/retrieve the knowledge object of the system. This operation includes two parts: searching relative concept and search/retrieve the knowledge. Let $N$ be the number of *concept tree* entries and $n$ be the number of *concept nodes* for each *concept tree*. The complexity of the first part is $O(\log N + \log n)$ because the concept tree entries are indexed according to a tree model. However, the number of concept entries as well as of concept nodes of a concept tree is negligible compared to the number of knowledge entries. So this complexity depends strongly on the cost of search/retrieve operations. Let $M$ be the number of meta-knowledge entries in the *KM core*, so the complexity of searching a meta-knowledge entry at this level is $O(\log M + C_s)$, where $C_s$ is the commu-nication cost between a node $s$ and the host node where the meta-knowledge repository is stored. This depends on the bandwidth between two nodes and the size of the data size. The complexity of retrieving a knowledge object is the same as for the search operation. However, the retrieve operation depends on the number of knowledge entries $m$ in the *local KM*.

Some tests have been launched to evaluate the search/retrieve performance. More details about these tests can be found in [48]. Next, we estimate the performance of the knowledge map architecture. Firstly, the structure of *concept tree* is based on the concept map [57], which is one of the advantages of this model. We can avoid the problem of semantic ambiguity as well as reduce the domain search to improve the speed and accuracy of the results. In the 1-n model (one server-n client nodes), the concept tree is implemented either only at the server node or at each client node. The client-server communication is needed when we interact with concept tree via the operations add, search, delete concept nodes or get the concept identity when adding new knowledge. In a large distributed system, this concept tree can be cached at each local node to reduce the communication cost because the number of operations of add/delete a concept node is very small compared to the number of search operations.

Secondly, the division of knowledge map into two main components (local and core) has some advantages: (i) the core component acts as a summary map of knowledge and it is a representation of knowledge about knowledge when combined with local *KM*; (ii) avoiding the problem of having the whole knowledge on one master node (or server), which is not feasible on very large distributed systems such as Grid. By representing knowledge meta-data by their relationship links, the goal is to provide an integration view of these knowledge.

Finally, this approach offers a knowledge map with flexible and dynamic architecture where users can easily update the *concept tree* repository as well as meta-knowledge entries. The current index technique used in a rule representative is an inverted list. However, we can improve it without affecting to whole system structure by using other index algorithms [53] or by applying compressed technique as discussed in [76]. Moreover, flexible and dynamic features are also reflected by mapping a knowledge to a *knowledge object*. The goal here is to provide a portable approach where knowledge object can be represented by different techniques such as an entity, an XML-based record, or a record of database, etc.

Another important discussion is the implementation of knowledge core. In the 1-n model, this component is implemented at the server node. However, as explained in the second paragraph of this section, the search operation time is higher than retrieve operation time because of the number of meta-knowledge entries. This will become a crucial bottleneck in a large-scale distributed system where only one server is dedicated to handle meta-knowledge of all other client nodes. An efficient solution is to use some nodes that can act as servers; i.e. each site handles the meta-knowledge of a group of client nodes. Nevertheless, which nodes will be chosen and under which criteria is not a straightforward task because one should satisfy the constraints related to the network topology, nodes' performance, etc.

## 6 Exploitation

### 6.1 Interface

The upper of the interface layer is a graphical user interface (GUI) (Fig. 14) allowing the development and the execution of DDM applications. By using this interface, users can build an *DDM* job including one or many tasks via building an executing flow chart. A task contains either one of the DM techniques such as classification, association rules, clustering, prediction or other data operations such as data pre-processing, data distribution. Firstly, users choose a tasks and then they browse and choose resources that are represented by graphic objects, such as computing nodes, datasets, DM tools and algorithms correspondent to DM technique chosen. These resources are either on local site or distributed on different heterogeneity sites with heterogeneous platforms. However, this system allows users to interact with them transparently at this level. The second step in the building of a DDM job is to establish links between tasks chosen, i.e. the execution order. By checking this order, this system can detect independent tasks that can be executed concurrently. Furthermore, users can also use this interface to publish new DM tools and algorithms. Besides, users can separately execute mining and preprocessing tasks provided (e.g for testing purposes) by choosing appropriate tools supplied by the system (Fig. 15).

This layer allows to visualise, represent as well as to evaluate results of an *DDM* application too. The discovered knowledges will be represented in many defined forms such as graphical, geometric, etc. This system supports different visualisation techniques which are applicable to data of certain types (discrete, continual, point, scalar or vector) and dimensions (1-D, 2-D, 3-D). It also supports the interactive visualisation which allows users to view the *DDM* results in different perspectives such as layers, levels of detail and help them to understand these results better. Besides the GUI, there are four modules in this layer: *DDM* task management, Data/Resource management, interpretation and evaluation.

The first module spans both Interface and Core layers of the system. The part in the Interface layer of this module is responsible for mapping user requirements via selected *DM* tasks and their resources to an executing schema of tasks correspondent. Another role of this part is to check the coherence between *DM* tasks of this executing schema for a given *DDM* job. The purpose of this checking is, as mentioned above, to detect independent tasks and then this schema is refined to obtain an optimal execution. After verifying the executing schema, this module stores it in a task repository that will be used

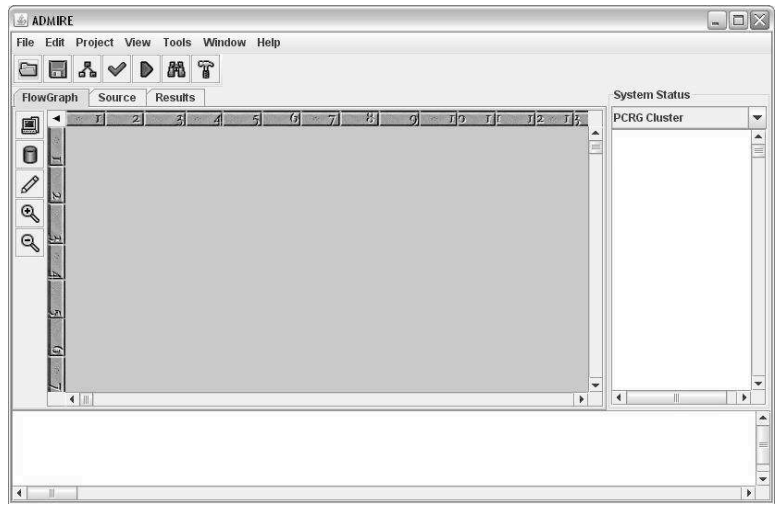

**Fig. 14.** *DDM* system inteface

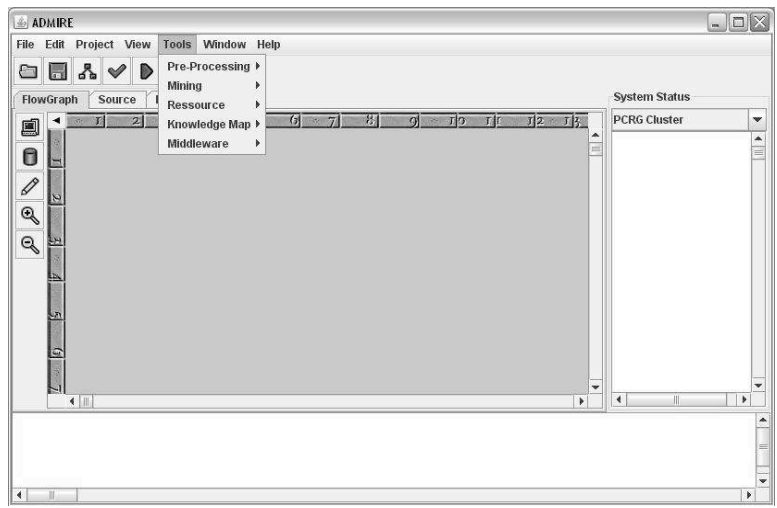

**Fig. 15.** System tools

by the lower part of this task management module in the core layer to execute this *DDM* job.

The second module allows to browse necessary resources in a set of re-sources proposed by system. This module manages the meta-data of all the available datasets and resources (computing nodes, *DM* algorithms and tools) published. The part in the Interface layer of this module is based on these meta-data that are stored in two repositories: datasets repository and re-sources repository to supply an appropriate set of resources depending on the given *DM* task. Data/Resource management module spans both Interface and Core layers of the system. The reason is that modules in the core layer also need to interact with data and resources to perform data mining tasks as well as integration tasks. In order to mask grid platform, data/resource management module is based on a data grid middleware, e.g. DGET[40].

The third module is for interpreting *DDM* results to different ordered presentation forms. Integrating/mining result models from knowledge map module in the core layer is explained and evaluated.

The last module deals with evaluation the *DDM* results by providing different evaluation techniques. Of course, measuring the effectiveness or usefulness of these results is not always straightforward. This module also allows experienced users to add new tools or techniques to evaluate knowledge mined. The last module deals with evaluation the *DDM* results by providing different evaluation techniques. Of course, measuring the effectiveness or usefulness of these results is not always straightforward. This module also allows experienced users to add new tools or techniques to evaluate knowledge mined.

## 6.2 An example of exploiting the *DDM* system via Knowledge Map

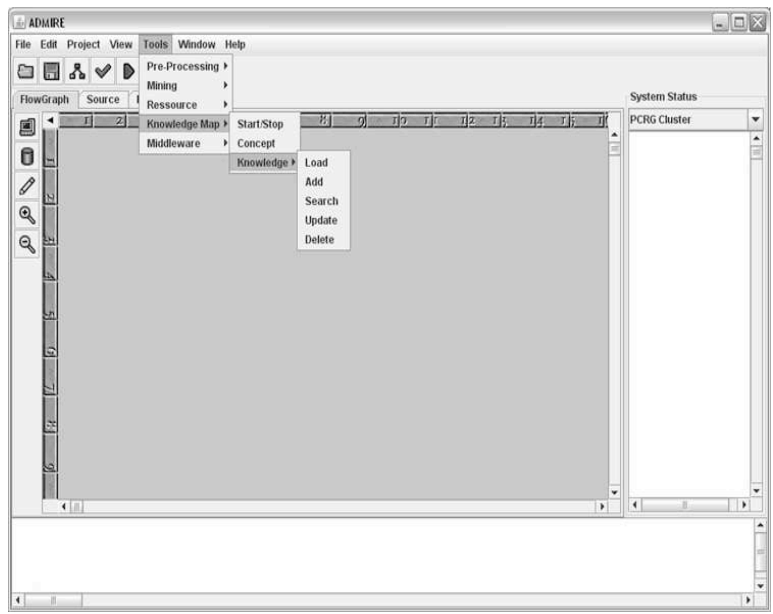

**Fig. 16.** Knowledge Map integrated in the *DDM* system.

*KM* (cf. section 5) is implemented and integrated in the *DDM* system (Fig.16). In this version, repositories of *KM core* and *Local KM* are in XML format. *KM* daemons are initially created by using the primitive ***"init"***. The primitive ***"stop"*** will terminate all the *KM Daemons*. A *KM* application can send requests to one or many remote sites.

As shown in Fig.13, for example, we first search all the meta-knowledge needed via primitive "find". The user can launch this task via either a command line or graphical user interface (Fig.17). We extract knowledge via primitive "retrieve". As an example with the system's GUI, Figure 18 shows a screenshot of meta-knowledge found in DBDC [42] clustering task. The user can decide which knowledge to be retrieved by selecting an appropriate row and then click on the retrieve button. In Figure 19 an example of a retrieved knowledge which is represented by a tree is shown. An important remark is that the users do not take into account the location of knowledge when searching/retrieving it.

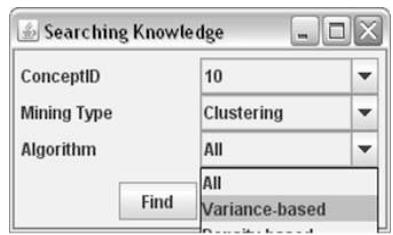

**Fig. 17.** Screenshot of searching Meta-knowledge.

| 🔲 111_Found.xml | | | | | |
|---|---|---|---|---|---|
| Global Knowle... | Algorithm | Data Type | # Instances | # Dimensions | Description |
| 9 | DBDC-local | Numerical | 161 | 2 | Using DBSCA... |
| 10 | DBDC-local | Numerical | 161 | 2 | Using DBSCA... |
|  |  | Retrieve | Exit |  |  |

**Fig. 18.** Screenshot of retrieving Knowledge basing on Meta-knowledge chosen.

Next, we present two scenarios of using knowledge map in the system. In the first one, the mined knowledge already exist at different sites of the system. In the second scenario, a distributed data mining task is executed on a system such as cluster or gird. In the first scenario, if the meta-knowledge of those mined knowledge have not been handled by the knowledge map, then the first step is to use knowledge map tool to create knowledge objects and store them in each site (*local KM*). Their *meta-knowledge* will be automatically submitted to the *meta-knowledge repository* at the *knowledge core map*. The users can also add more appropriate concepts to their knowledge. This step needs more interactions with the user. The user, then, can exploit these meta-knowledge and knowledge object in their integration process or only explore the knowledge. In the second scenario, after the local mining processes have been completed, the local mined knowledge is built in each site. Its meta-knowledge is created and stored in local repository. Then, the integration process uses these meta-knowledge to retrieve information required.

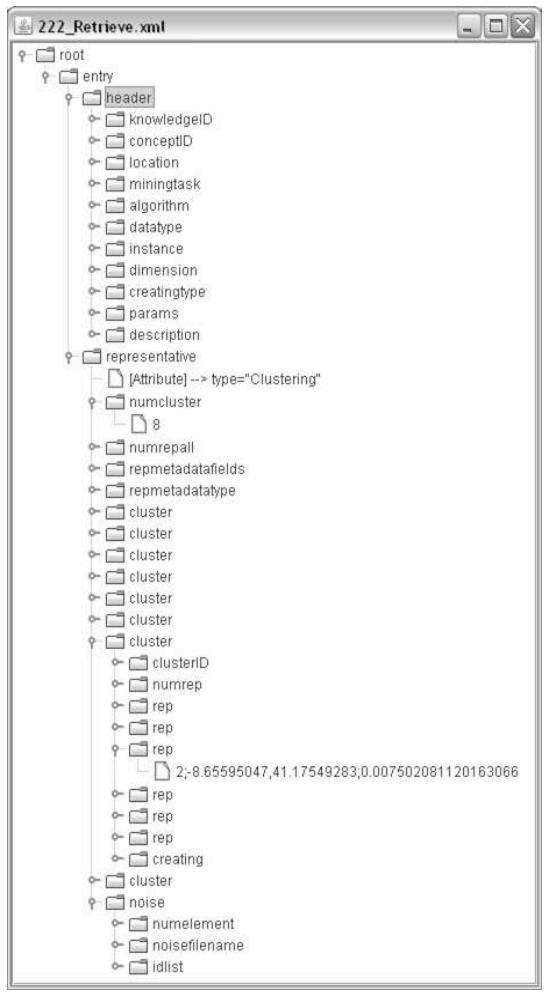

**Fig. 19.** Knowledge retrieved, tree format.

# 7 Related Works of DDM frameworks on Grid platforms

We present three most recent *DDM* projects for heterogeneous data and platforms: Knowledge Grid [15] [16], Grid Miner [9] [10] and Discovery Net [22]. The first two projects use Globus Toolkit [36] as a Grid middleware.

## 7.1 Knowledge Grid

Knowledge Grid (KG) is a framework for implementing distributed knowledge discovery. This framework aims to deal with multi-owned, heterogeneous data. This project is developed by Cannataro el al. at University "Magna Graecia" of Catanzaro, Italy. The architecture of KG is composed of two layers: Core K-Grid and High level K-Grid.

The first layer includes Knowledge Directory service (KDS), Resource allocation and execution management service (RAEMS). KDS manages metadata: data sources, data mining (DM) software, results of computation, etc. that are saved in KDS repositories. There are three kinds of repositories: Knowledge Metadata Repository for storing data, software tool, coded information in XML; Knowledge Base Repository that stores all information about the knowledge discovered after parallel and distributed knowledge discovery (PDKD) computation. Knowledge Execution Plan Repository stores execution plans describing PDKD applications over the grid. RAEMS attempts to map an execution plan to available resource on the grid. This mapping must satisfy users, data and algorithms requirements as well as their constraints.

High level layer supplies four service groups used to build and execute PDKD computations: Data Access Service (DAS), Tools and Algorithms Access Service (TASS), Execution Plan Management Service (EPMS) and Results Presentation Service (RPS). The first service group is used for the search, selection, extraction, transformation, and delivery of data. The second one deals with the search, selection, download DM tools and algorithms. Generating a set of different possible execution plans is the responsible of the third group. The last one allows to generate, present and visualize the PDKD results.

The advantage of KG framework that it supports distributed data analysis and knowledge discovery and knowledge management services by integrating and completing the data grid services. However this approach only concerns the distributed architecture but not *DDM* algorithms. Besides, KG does not provide a management of knowledge metadata in their relationships to support the integration view of the knowledge as well as the coordination of different local mining processes. There is moreover no distinct separation in between resource, data, and knowledge.

## 7.2 GridMiner

GridMiner is an infrastructure for distributed data mining and data integration in Grid environments. This infrastructure is developed at Institute for Software Science, University of Vienna. GridMiner is a OGSA-based data mining approach. In this approach, distributed heterogeneous data must be integrated and mediated by using OGSA-DAI [24] before passing data mining phase. Therefore, they have divided data distribution in four data sources scenarios: single, federate horizontal partitioning, federate vertical partitioning and federate heterogeneity.

The structure of GridMiner consists of some elements: Service Factory (GMSF) for creating and managing services; Service Registry (GMSR) that is based on standard OGSA registry service; DataMining Service (GMDMS) that provides a set of data mining, data analysis algorithms; PreProcessing Service (GMPPS) for data cleaning, integration, handling missing data, etc.; Presentation Service (GMPRS) and Orchestration Service (GMOrchS) for handling complex and long-running jobs.

The advantages of GridMiner is that it is an integration of Data Mining and Grid computing. Moreover, it can take advances from OGSA. However, it depends on Globus ToolKit as well as OGSA-DAI for controlling data mining and other activities across Grid platforms. Besides, distributed heterogeneous data must be integrated before the mining process. This approach is not appropriate for complex heterogeneous scenarios.

### 7.3 Discovery Net

Discovery Net project proposes an architecture to support the knowledge discovery process on Grid platforms.

Discovery Net (*DN*) architecture is composed of three main components: Knowledge Servers, Resource Discovery Server and Meta-information Server. The first component is a warehouse of information about discovery process performed by *DN*. It provides three main functions: storage service, reporting service and application generation service. The Resource Discovery Server component is a registry server which is used to deploy and map services to computational resources for execution. The last component is responsible for the management of data types used by services in the system.

A remarkable pros of this architecture is that it is based on the service model concept and can be used on any service-based Grid platforms. However, as Knowledge Grid resumed above, this approach does not concerns *DDM* algorithms.

## 8 Conclusion

In this chapter, we have presented a *DDM* system based on Grid environments to execute new distributed data mining techniques on very large and distributed heterogeneous datasets. The architecture and motivation for the design have been presented. We have also discussed some related projects and compared them with our approach. We have developed prototypes for each layer of the system to evaluate the system features, test each layer as well as whole framework and building simulation and *DDM* test suites.

Knowledge map layer, key layer of this system, is integrated in this framework. Experimental results on real-world applications are also produced [28] and allow us to test and evaluate the system robustness and the distributed data mining approaches at very large scale. Throughout estimations of each component and its functionality, we can conclude that knowledge map is an efficient system in a large-scale and distributed environment. It satisfies the needs for managing, exploring, and retrieving the mined knowledge of *DDM* in large distributed environment. We are currently working on the knowledge map structure that takes in account the network and the distributed system (such as Grid) features. These features include the mapping of the knowledge map onto a virtual network topology such as *TreeP*.

*DDM* algorithms presented in the section 4 above are also being integrated in the system in the module of distributed data mining techniques. The preliminary evaluation of our approach shows that it is efficient and flexible.

In the future works, more *DDM* algorithms will be developed and integrated in this system. Besides, a workflow engine will also be built on the top of the system in order to conduct efficient implementations on the Grid at both the application level and the knowledge management level.